\documentclass[aps,showpacs,showkeys,preprint]{revtex4}
\usepackage{latexsym,amsmath}

\begin{document}


\title{Entropy of Reissner-Nordstrom Black Holes with Minimal Length Revisited}

\author{Myungseok Yoon}
  \email{younms@sogang.ac.kr}
  \affiliation{Center for Quantum Spacetime, Sogang
    University, Seoul 121-742, Korea}
\author{Jihye Ha}
  \affiliation{Department of Physics, Sogang University, Seoul
    121-742, Korea}
\author{Wontae Kim}
  \email{wtkim@sogang.ac.kr}
  \affiliation{Department of Physics and Center for Quantum Spacetime,
    Sogang University, Seoul 121-742, Korea}

\date{\today}

\begin{abstract}
Based on the generalized uncertainty principle, we study the entropy
of a four-dimensional black hole by counting degrees of freedom
near the horizon and obtain the (finite) entropy proportional to the surface
area at the horizon without a cutoff introduced in the conventional
brick-wall method.
\end{abstract}


\maketitle


Bekenstein suggested that the entropy of a black hole is proportional
to the surface area at the event horizon \cite{bekenstein}, and
subsequently Hawking has proved that based on the quantum field theory
the entropy of the Schwarzschild black hole satisfies the area law
\cite{hawking}. Motivated by these works, there have been many
interesting works to study the statistical origin of the entropy
\cite{zt,fn,cw,mann}. One of them is the well-known brick-wall method
\cite{thooft}, in which the divergence near the horizon can be removed
by introducing a brick-wall cutoff. It can help us to understand the
origin of entropy in various black holes
\cite{su,dlm,hk,mukohyama}. However, the introduction to the brick
wall cutoff seems to be more or less unnatural. Recent calculations of
the free energy and the entropy have been studied 
using the generalized uncertainty relation, in which there exists the
minimal length \cite{li,liu,lhz,sl,kkp}. In this regime, the bound of
the entropy has been obtained in Ref.~\cite{li,liu,lhz} and some
detail calculations for the specific models are done in
Ref.~\cite{sl,kkp}. The simplest way to generalize the uncertainty
relation is to promote it to $ \Delta x \Delta p \ge
\hbar + \frac{\lambda}{\hbar} (\Delta p)^2$, which shows that there
exists a minimal length, $\Delta x \ge 2 \sqrt{\lambda}$ \cite{garay,kmm}.

Specifically, as for a four-dimensional spherically symmetric black hole
\cite{sl,kkp}, the detail higher-order correction to the entropy near
the horizon $r_0$ can be calculated as
\begin{equation}
  S = \frac{r_0^2}{12\lambda} \left[ 1 - \frac{4\pi^2}{3} + 64
      \zeta(3) - \frac{16\pi^4}{5} - \frac{1024\zeta(5)}{3} +
      O((4\pi)^3) \right].
\end{equation}
Note that the higher-order corrections are not negligible, more worse,
it is not convergent so that the result may not be reliable.
So, in this paper, we would like to obtain the finite convergent
entropy of the Reissner-Nordstrom(RN) black hole near the horizon 
with the generalized uncertainty principle(GUP) in the regime of the
brick-wall method. There is no ultraviolet divergence at all as long
as we use the generalized uncertainty relation when we count the
density of states even at the horizon.

Now, we consider the RN black hole, whose metric is given by
\begin{equation}
  \label{metric}
  ds^2 = - f(r) dt^2 + \frac{dr^2}{f(r)} + r^2 d\theta^2 + r^2
  \sin^2\theta d\varphi^2,
\end{equation}
with $f(r) = 1 - 2M/r + Q^2/r^2$, where $M$ and $Q$ are the mass and
the electric charge of the black hole, respectively. Then, the inner
($r_-$) and the outer ($r_+$) horizons are obtained as $ r_\pm = M \pm
\sqrt{ M^2 - Q^2}$. In this black hole, the Hawking Temperature is $T
= \beta^{-1} = \kappa/(2\pi) = (r_+ - r_-)/(4\pi r_+^2)$, where
$\kappa$ is the surface gravity at the outer horizon.

The Klein-Gordon equation for a scalar field on this background
becomes
\begin{equation}
  \label{KG}
  (\Box-\mu^2) \Phi = 0,
\end{equation}
where $\mu$ is the mass of the field. Substituting the wave function
$\Phi = \exp(-i\omega t) \psi(r,\theta,\varphi)$ into Eq.\
(\ref{KG}), we obtain
\begin{equation}
  \label{KG:r}
  \frac{\partial^2 \psi}{\partial r^2} + \left( \frac{2}{r} +
    \frac{1}{f} \frac{df}{dr} \right) \frac{\partial\psi}{\partial r}
  + \frac{1}{f} \left[ \frac{\omega^2}{f} - \mu^2+ \frac{1}{r^2} \left( 
      \frac{\partial^2}{\partial\theta^2} + \cot\theta
      \frac{\partial}{\partial \theta} +
      \csc^2\theta\frac{\partial^2}{\partial\varphi^2}
    \right) \right]\psi = 0.
\end{equation}
By using the WKB approximation with $\psi \sim \exp[iS(r,\theta,
\varphi)]$, we obtain $p_r^2 = f^{-1} \left[ \omega^2/f - \mu^2 -
r^{-2} p_\theta^2 - r^{-2} \csc^2\theta \, p_\varphi^2 \right], $
where $p_r = \partial S / \partial r$, $p_\theta = \partial S /
\partial \theta$, and $p_\varphi = \partial S / \partial \varphi$.
And the square module of momentum is given by
$p^2 = p_i p^i = g^{rr} p_r^2 + g^{\theta\theta} p_\theta^2 +
  g^{\varphi\varphi} p_\varphi^2 =\omega^2/f - \mu^2$.
Then, the number of quantum states is
obtained as
\begin{eqnarray}
  n(\omega) &=& \frac{1}{(2\pi)^3} \int dr d\theta d\varphi dp_r
    dp_\theta dp_\varphi \frac{1}{(1+\lambda p^2)^3} \nonumber \\
    &=& \frac{2}{3\pi} \int dr \frac{r^2 (\omega^2/f - %
      \mu^2)^{3/2}}{\sqrt{f}[1 + \lambda (\omega^2/f-\mu^2)]^3},
    \label{n}
\end{eqnarray}
where $\omega \ge \mu\sqrt{f}$. Note that it is finite even at the
horizon. This is the reason why the brick-wall cutoff is not necessary
in our calculation. The free energy for a scalar field is given by 
$  F = - \int_{\mu\sqrt{f}}^\infty d\omega
  \frac{n(\omega)}{e^{\beta\omega} -1}$,
and the entropy is obtained as
\begin{eqnarray}
  S &=& \beta^2 \frac{\partial F}{\partial\beta} \nonumber \\
  &=& \frac{2\beta^2}{3\pi} \int dr \frac{r^2}{\sqrt{f}}
    \int_{\mu\sqrt{f}}^\infty d\omega \frac{\omega(\omega^2/f -
    \mu^2)^{3/2}}{4 \sinh^2 \frac12\beta\omega\, [1 +
    \lambda(\omega^2/f - \mu^2)]^3} \nonumber \\
  &=& \frac{\beta^3}{12\pi\lambda^3} \int dr\, r^2 f \int_{x_0}^\infty
    dx \frac{x(x^2 - x_0^2)^{3/2}}{\sinh^2 x\, [x^2 - x_0^2 + \beta^2
    f/(4\lambda)]^3}, \label{S}
\end{eqnarray}
where $x \equiv \frac12 \beta\omega$ and $x_0 = \frac12
\beta\mu\sqrt{f}$. 

We now consider a thin layer between $r_+$ and $r_+ + \epsilon$ for a
nonextremal RN black hole near the horizon for the mode counting since
the density of state is dominant in this region.
Moreover, since $x_0$ goes to zero near the horizon, the
entropy becomes 
\begin{equation}
  \label{S:massless}
  S = \frac{\beta^3}{12\pi\lambda^3} \int_0^\infty dx\,
  \frac{x^4}{\sinh^2 x} I(x),
\end{equation}
where
\begin{equation}
  \label{I:def}
  I(x) = \int_{r_+}^{r_+ + \epsilon} dr\, \frac{r^2 f}{[x^2 + \beta^2
    f/(4\lambda)]^3}.
\end{equation}
Near the horizon, the metric can be simply written as $f(r)
\approx f'(r_+)(r - r_+)$, where $f'(r_+) = 2\kappa = (r_+ -
r_-)/r_+^2$. From the metric (\ref{metric}), the minimal length is
obtained as 
\begin{equation}
  \label{lambda}
  2 \sqrt{\lambda} \equiv \int_{r_+}^{r_+ + \epsilon}
  \frac{dr}{\sqrt{f}} = \sqrt{\frac{2\epsilon}{\kappa}}.
\end{equation}
Then, the integral (\ref{I:def}) is calculated as
\begin{equation}
  \label{I}
  I(x) \approx \frac{r_+^2 f'(r_+) \epsilon^2}{4x^2 [x^2 + \beta^2 f'(r_+)
    \epsilon/(4\lambda)]^2} = \frac{2\kappa^3 \lambda^2 r_+^2}{x^2 +
    4\pi^2}.
\end{equation}
Substituting Eq.\ (\ref{I}) into Eq.\ (\ref{S:massless}), we
obtain
\begin{equation}
  \label{S:int}
  S = \frac{4\pi^2 r_+^2}{3\lambda} \int_0^\infty dx\,
  \frac{x^2}{\sinh^2 x(x^2 + 4\pi^2)^2}.
\end{equation}
Then, the integrand in Eq.\ (\ref{S:int}) can be 
regarded as a complex function $h(z) \equiv z^2 / [\sinh^2 z (z^2 +
4\pi^2)^2]$ and poles of $h(z)$ are given by $z=in\pi$, where
$n$'s are integers except zero. The residues of $h(z)$ are $i/(24\pi)$ at
$z=2i\pi$ and $-2in(n^2+4)/[\pi^3(n^2-4)^3]$ at $z=in\pi$ for $n \ne
2$. By the residue theorem, the entropy becomes
\begin{equation}
  \label{S:final}
  S = \frac{4\pi^2 r_+^2}{3\lambda} \left( -\frac{1}{24} +
      \frac{2}{\pi^2} \sum_{\substack{n = 1 \\ n \ne 2}}^\infty
      \frac{n(n^2 +4)}{(n^2 - 4)^3} \right) 
    = \frac{4\pi r_+^2}{3\lambda} \left(-
      \frac{\pi}{24} - \frac{25}{32\pi} + \frac{\zeta(3)}{\pi}
      \right),
\end{equation}
where this finite result is exact in the near horizon limit. 
This entropy can be written as $S=\frac14 A \cdot
\frac{\delta}{\lambda}$, where $A=4\pi r_+^2$ and $\delta \equiv
\frac13 [\frac{4}{\pi}\zeta(3) - \frac{25}{8\pi} - \frac{\pi}{6}]$.
In order to satisfy the area law of entropy for this black hole, $\lambda$
is required to be the same with $\delta$. By recovering the dimension, the
minimal length is obtained as $2\sqrt{\lambda} = 2\sqrt{\delta}\, \ell_p
\approx 0.127484\times \ell_p$, where $\ell_p$ is the Plank length.

In conclusion, our calculation shows that it is possible to obtain the
positive definite convergent entropy in the near horizon limit of the
RN black hole by using the brick wall method along with the
generalized uncertainty principle. Conversely, the requirement of the
area law determines the arbitrary positive constant $\lambda$ in the
generalized uncertainty principle. And, the most interesting thing to
distinguish from the conventional method is that there is no
divergence at all, which seems to be a generic behavior as long as we
use the generalized uncertainty relation.


\section*{Acknowledgments}
This work was supported by the Science Research Center Program of the
Korea Science and Engineering Foundation through the Center for
Quantum Spacetime (CQUeST) of Sogang University with grant number
R11-2005-021.


\end{document}